\documentclass[11pt]{article}
\usepackage{moriond,epsfig}

\def\Journal#1#2#3#4{{#1} {\bf #2}, #3 (#4)}

\def\NPB{{\em Nucl. Phys.} B}
\def\PLB{{\em Phys. Lett.}  B}
\def\PRL{\em Phys. Rev. Lett.}
\def\PRD{{\em Phys. Rev.} D}

\def\be{\begin{equation}}
\def\ee{\end{equation}}
\def\bea{\begin{eqnarray}}
\def\eea{\end{eqnarray}}

\begin{document}
\vspace*{4cm}
\title{SMALL-{\bf $x$} PHYSICS AND THE DETECTION OF UHE NEUTRINOS}

\author{N. Armesto, C. Merino, G. Parente, and
E. Zas}

\address{Departamento de F\'\i sica de Part\'\i culas, Facultade de 
F\'\i sica, \\ 
and Instituto Galego de F\'\i sica de Altas Enerx\'\i as (IGFAE), \\ 
Universidade de Santiago de Compostela, Galiza, Spain}

\maketitle\abstracts{We evaluate both the tau lepton energy loss produced by photonuclear
interactions and the neutrino charged current cross section
at ultra-high energies, both
relevant to neutrino bounds with Earth-skimming tau neutrinos.}

Neutrino bounds at ultra-high energies (UHE) have been successfully 
stablished~\cite{AugerPRL} by the
Pierre Auger Collaboration by looking for tau neutrinos that reach and exit
the Earth. This Earth-skimming channel
directly depends both on the tau  range (the energy loss)
and on the neutrino charged current cross section
which determine the amount of matter with which the neutrino has to interact 
to produce an emerging tau. Contrary to the muon case,
for tau leptons of energies above $E = 10^7$ GeV
photonuclear interactions are responsible
for the largest and the most uncertain contribution~\cite{Dutta2001}.

The $Q^2$ scale that contributes to the tau energy loss 
is low and moderate $Q^2$ at very low $x$,
where perturbative and non perturbative QCD
effects are mixed. The charged current (CC) neutrino cross section is produced by $W$-boson
exchange what sets the relevant scale of $Q^2$ to
values up to $M_W^2~$ at low $x$, a region
where perturbative QCD is expected to work.
In both cases the relevant $x$ range
lies well outside the regions where structure functions are
measured, so one has to rely on extrapolations which contain significant
uncertainties.

We present the combined analysis of both the tau energy loss
and of the neutrino-nucleus cross section~\cite{paper,nosoICRC07}. 
Because of the large uncertainties in the existing models, the fact that
none of them simultaneously covers the kinematical region relevant for
both the tau energy loss and the neutrino-nucleus cross section, 
and the need of consistency in both calculations, we use
and extend available models with the aim of estimating the theoretical
uncertainty by considering extreme results.
In this way, in the frame of the most relevant models, we scan the
range of possible scenarios for the extrapolation of
structure functions to the relevant $x$ and $Q^2$ range.

The contribution to the fractional average energy loss per unit depth
of taus from photonuclear
interactions above the scale of a few TeV,
\begin{eqnarray}
b(E) = -\frac{1}{E} \left<\frac{dE}{dX}\right> \; ,
\end{eqnarray}
is obtained by integration of the
lepton-nucleus differential cross section, $d\sigma^{lA}/dy$:
\begin{eqnarray}
b(E) =
\frac{N_A}{A} \int dy \; y \int dQ^2 \frac{d\sigma^{lA}}{dQ^2 dy} \; ,
\end{eqnarray}
where $N_A$ is Avogadro's number, $A$ the mass number,
and $y$ the fraction of energy
lost by the lepton in the interaction. For the lepton-nucleus differential 
cross section we consider the general expression for virtual photon 
exchange in terms of structure functions~\cite{paper}.

\begin{figure}[htb]
\centering
\includegraphics[width=.55\hsize]{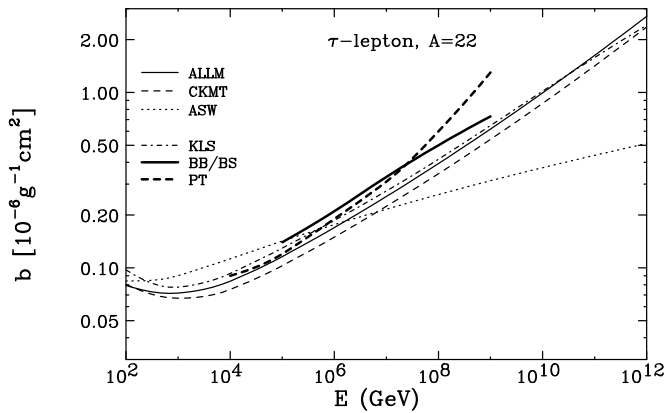}
\vskip -0.3cm
\caption{\footnotesize The photonuclear energy loss rate, $b(E)$, computed in different
models.}
\label{figloss}
\end{figure}

The photonuclear contributions to $b(E)$ computed
(for standard rock $A=22$ throughout all this paper)
with ALLM~\cite{ALLM} and with CKMT~\cite{CKMT} structure functions, and the same nuclear
corrections~\cite{Dutta2001}, give very close results
(see Fig.~\ref{figloss}). The BB/BS calculation gives~\cite{BB,BS2003}
the largest of the predicted energy loss rates up to energies of the order
$E=10^7$~GeV. The BB/BS, ALLM, and CKMT calculations of the photonuclear contribution
to tau energy loss, $b(E)$, agree within a 30~$\%$ and go
approximately parallel for all energies, which is an indication of a
systematic normalization difference of the structure functions in each
model. The lowest values of $b(E)$ at high energies is
obtained with the ASW~\cite{ASW} structure functions,
which are based on the geometric scaling property that
all data on $\sigma^{\gamma^* p}$ and on
$\sigma^{\gamma^* A}$ lie on a single universal curve
in terms of the scaling variable $\tau=Q^2/Q^2_{sat}$
whose form is inspired in saturation physics.
Above the scale of $E=10^7$~GeV the PT result exceeds~\cite{Petrukhin} all other
existing predictions by at least a factor 2 already at $E=10^9$~GeV
(i.e. a factor 4 with respect the ASW prediction, see Fig.~\ref{figloss}).
Thus the PT prediction can be considered as an estimate of the upper limit
of the tau energy loss at UHE.

\begin{figure}[htb]
\centering
\vspace{-0.2cm}
\includegraphics[width=.55\hsize]{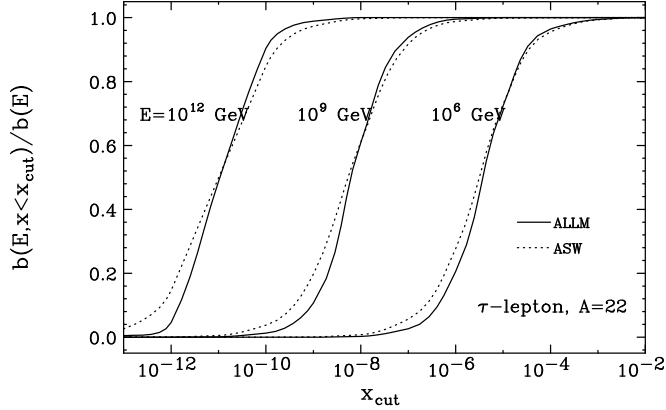}
\vskip -0.3cm
\caption{\footnotesize The relative contribution of $x<x_{cut}$
to the photonuclear energy loss rate, $b(E)$.}
\label{figrelx}
\vskip -0.1cm
\end{figure}

Much of the uncertainty in the
tau energy loss is actually due to nuclear effects.
The choice of nuclear corrections~\cite{Dutta2001,ASW,BM2002}
translates into differences in the calculated value of $b(E)$
by a factor rising from 1.5 to 2.5
as the tau energy increases in the range $E=10^{6}$-$10^{9}$~GeV
when using the ALLM  structure function.
This energy range corresponds to the region
of very low $x$ where differences in the nuclear correction factor are large~\cite{paper}.
In Fig.~\ref{figrelx} it is shown how the small $x$ contribution becomes more and more
important as energy increases.

We have also studied how the uncertainties in the
$F_2$ structure function at low $x$
affect the CC neutrino deep inelastic cross section
that is expressed in terms of the
structure function $F_2$ as follows:
\begin{eqnarray}
\frac{d\sigma_{CC}^{\nu N}}{d Q^2 dy} =
\frac{G_F^2}{4 \pi} \left(\frac{M_W^2}{M_W^2+Q^2}\right)^2
\frac{F_2^{\nu N}}{y} [ 1+(1-y)^2] \; ,
\end{eqnarray}
where $E$ is the neutrino energy and $y$ the fraction of energy
lost by the neutrino in the interaction. In this expression
$F_L$ and $xF_3$ contributions are neglected since $F_L$ tends to zero
as $Q^2$ rises and $xF_3$ deals basically with the valence partons
which hardly contribute at the low $x$ values relevant for
the cross section.

The $F_2$ structure function for neutrino interaction is related
to the $F_2$ structure function
for charged lepton interactions by the ratio of the weak and electromagnetic
couplings through $F_2^{\nu N}=18/5 \; F_2^{lN}$ (assuming a symmetric sea).
Then, to calculate the neutrino-nucleon cross section at
high energies, the structure function $F_2$ for charged lepton
interaction valid up to very low $x$ and high $Q^2$ must be used.

The neutrino-nucleon cross sections from ALLM and CKMT structure functions
are clearly
below predictions from  modern parton densities~\cite{Anchordoqui}, so
to discuss
the theoretical uncertainties in the estimation of
the CC neutrino-nucleon cross section we have taken
the parameterization of $F_2$
{\it \`a} $la$ BCDMS obtained by the SMC Collaboration~\cite{SMC}, which
correctly represents the existing
experimental data at high $Q^2$ and provides
a smooth connection at neutrino energies around $E=10^7$~GeV
with the parton density prediction.

We have performed three different extrapolations at low $x$ of the
$F_2$ parameterization {\it \`a} $la$ BCDMS,
one following the ASW structure function, a second one from the
phenomenological parameterization fitting
low $x$ HERA data \cite{HERA}, and the third one which corresponds to
the double logarithmic approximation (DLA) in QCD \cite{KOPA} 
(KOPA). The ASW and KOPA structure functions are valid
at low $x$, $x<0.01$, i.e. at high energies.

\begin{figure}[htb]
\centering
\vspace{-0.2cm}
\includegraphics[width=.55\hsize]{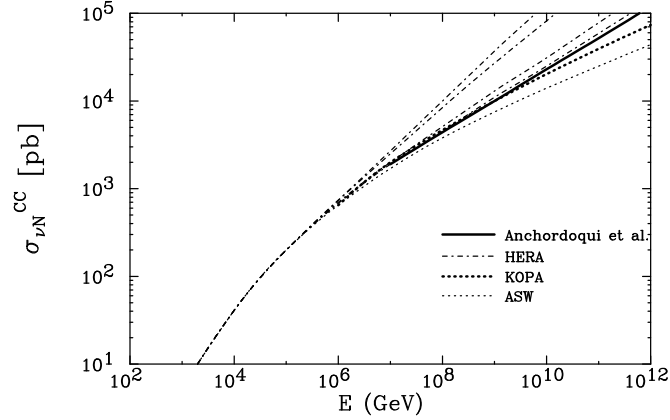}
\vskip -0.3cm
\caption{\footnotesize The neutrino-nucleon CC cross section as a function of the
neutrino energy,~$E$.}
\label{fignuxsection2}
\end{figure}

We can see in Fig.~\ref{fignuxsection2} 
that in comparison with the prediction obtained with evolved QCD parton
densities~\cite{Anchordoqui}, both KOPA
and ASW
estimations are below at high energies.
On the other hand the extrapolation of the HERA based parameterization
with the exponent
$\lambda=0.0481 \ln(Q^2/0.292^2)$ ($F_2 \sim x^{-\lambda}$),
produces an extremely fast increase of the cross
section with energy, since this exponent rises to values above $\lambda \sim 0.5$
when $Q^2$ becomes large, in contradition with
perturbative calculations.
For the more realistic scenarios, when the rise of the exponent freezes to
smaller values $\lambda < 0.4$, our prediction supports the result
obtained in previous detailed analysis~\cite{Anchordoqui}.
When considering only physically motivated extrapolations,
the theoretical uncertainty at $E=10^{9}$ GeV is a factor 2.
In the case of the CC neutrino-nucleon cross section the importance of nuclear effects
at high energies is expected to be small~\cite{CastroPena}.

The detection of UHE $\tau$-neutrinos is dominated by small $x$ physics both at low and high $Q^2$.
The effects discussed in this paper should be accounted for in the proccess of optimizing
the determination of a bound from Earth-skimming $\nu_{\tau}$. 

\section*{Acknowledgments}
This paper was supported by Ministerio de Educaci\'on y Ciencia of Spain under 
the Spanish Consolider-Ingenio 2010 Programme CPAN (CSD2007-00042) and projects 
FPA2005-01963 and FPA2004-01198, and by Xunta de Galicia under grant 2005 PXIC20604PN 
and Conseller\'\i a de Educaci\'on.
N.A. acknowledges financial support by Ministerio de Educaci\'on y Ciencia
(MEC) of Spain under a Ram\'on y Cajal contract.

\end{document}